\documentclass{aastex}
\usepackage{spr-astr-addons}
\usepackage{url}

\RequirePackage{color}

\begin{document}

\title{Light curve solutions of the eclipsing eccentric binaries KIC 8111622,
KIC 10518735, KIC 8196180 and their out-of-eclipse variability}
\slugcomment{Not to appear in Nonlearned J., 45.}
\shorttitle{KIC 8111622, KIC 10518735 and KIC 8196180}
\shortauthors{D. Kjurkchieva and D. Vasileva}

\author{Diana P. Kjurkchieva\altaffilmark{1}} \and \author{Doroteya L. Vasileva\altaffilmark{1}}
\affil{Department of Physics and Astronomy, Shumen University, 9700 Shumen, Bulgaria}

\begin{abstract}
We determined the orbits and stellar parameters of three eccentric
eclipsing binaries by light curve solutions of their \emph{Kepler}
data. KIC 8111622 and KIC 10518735 undergo total eclipses while
KIC 8196180 reveals partial eclipses. The target components are G
and K stars, excluding the primary of KIC 8196180 which is early F
star. KIC 8196180 reveals well-visible tidally-induced feature at
periastron, i.e. it is an eclipsing heartbeat star. The
characteristics of the observed periastron feature (shape, width
and amplitude) confirm the theoretical predictions. There are
additional out-of-eclipse variations of KIC 8196180 with the
orbital period which may be explained by spot activity of
synchronously rotating component. Besides worse visible periastron
feature KIC 811162 exhibits small-amplitude light variations whose
period is around 2.3 times shorter than the orbital one. These
oscillations were attributed to spot(s) on asynchronously rotating
component.
\end{abstract}


\section{INTRODUCTION}

The eccentric binaries are important objects for the modern
astrophysics because they present probes for study of tidal
interaction. The gradual loss of energy due to tidal forces leads
to circularization of the orbit and the synchronization of the
rotation of these stars with their orbital motion \citep{Tas88,
Tas92, Zahn89, Regos05}.

The unprecedented high-accuracy of \emph{Kepler} observations and
their continuity \citep{bor10, koch10} allowed to discover many
new eccentric binaries as well as to confirm the second-order
effects predicted by \cite{Kumar95}: light feature at the
periastron and tidally-excited oscillations \citep{Welsh11, ful11,
bur11, Th12, nic12, ham13a}. The newly discovered objects were
called ''heartbeat'' (HB) stars. They provide important tests for
the stellar astrophysics and information about the stellar
interiors \citep{ham13b}.

The number of heartbeat stars found in the \emph{Kepler} data
gradually increases \citep{Kirk16}. Some of them have been objects
of follow-up spectroscopy which shows a good agreement between the
spectroscopic and photometric orbital elements \citep{Smullen15,
Shporer16}. Last years we studied around 30 eccentric
\emph{Kepler} binaries, 12 of them turned out heartbeat stars
\citep{kjur15p, kjur16na, kjur16apss, kjur16aj}.

The goal of this study is determination of the orbits and physical
parameters of three eccentric binaries, KIC 8111622, KIC 10518735,
KIC 8196180, as well as investigation of their out-of-eclipse
variability to search for tidally-excited effects. Table 1
presents available information for these targets from the EB
catalog \citep{Prsa11, Slawson11, Kirk16}.

\begin{table*}
\caption{Parameters of the targets from the EB catalog: orbital
period \emph{P}, mean temperature $T_{m}$, widths $w_{1,2}$ of the
eclipses (in phase units), depths of the eclipses $d_{1,2}$ (in
flux units), phase of the secondary eclipse $\varphi_{2}$}
\label{tab:1} \centering
\begin{scriptsize}
\begin{tabular}{ccccccccc}
\hline\hline
Kepler ID& $P$ [d] & $K_{m}$ & $T_{m}$ [K]& $w_1$ & $w_2$ & $d_1$ & $d_2$ & $\varphi_{2}$  \\
\hline
8111622  & 15.4460592 & 15.488 & 5603 & 0.0104 & 0.0135 & 0.318 & 0.028 & 0.183 \\
10518735 & 19.515     & 16.739 & 5146 & 0.0084 & 0.0093 & 0.208 & 0.142 & 0.815 \\
8196180  &  3.6716611 & 12.814 & 7114 & 0.0409 & 0.0409 & 0.128 & 0.041 & 0.404\\
\hline\hline
\end{tabular}
\end{scriptsize}
\end{table*}

\begin{figure}
   \centering
   \includegraphics[width=7.5cm, angle=0]{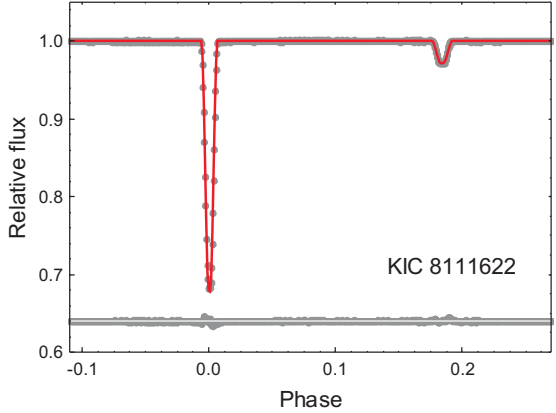}
    \caption{Top panel: the binned folded \emph{Kepler} data of KIC 8111622
    (black points) with their \textsc{PHOEBE} fit (red line);
    Bottom panel: the corresponding residuals from the model subtracted
    data for all phases (including eclipses)}
   \label{Fig1}
   \end{figure}

\begin{figure}
   \centering
   \includegraphics[width=7.5cm, angle=0]{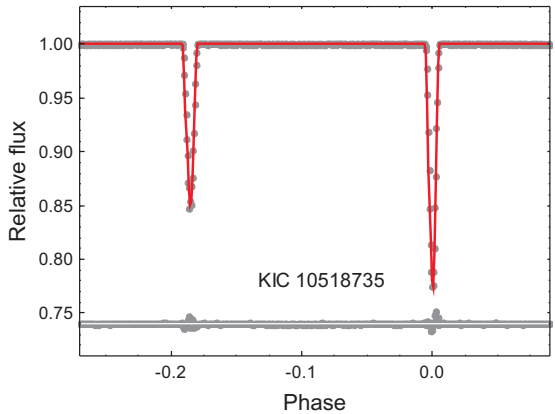}
    \caption{The same as figure 1 for KIC 10518735}
   \label{Fig2}
   \end{figure}

\begin{figure}
   \centering
   \includegraphics[width=7.5cm, angle=0]{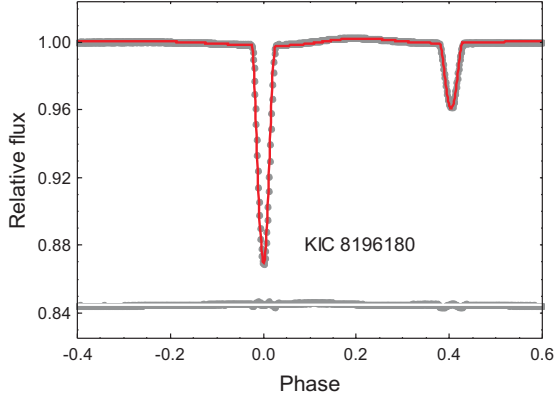}
    \caption{The same as figure 1 for KIC 8196180}
   \label{Fig3}
   \end{figure}

\begin{figure}
   \centering
   \includegraphics[width=8cm, angle=0]{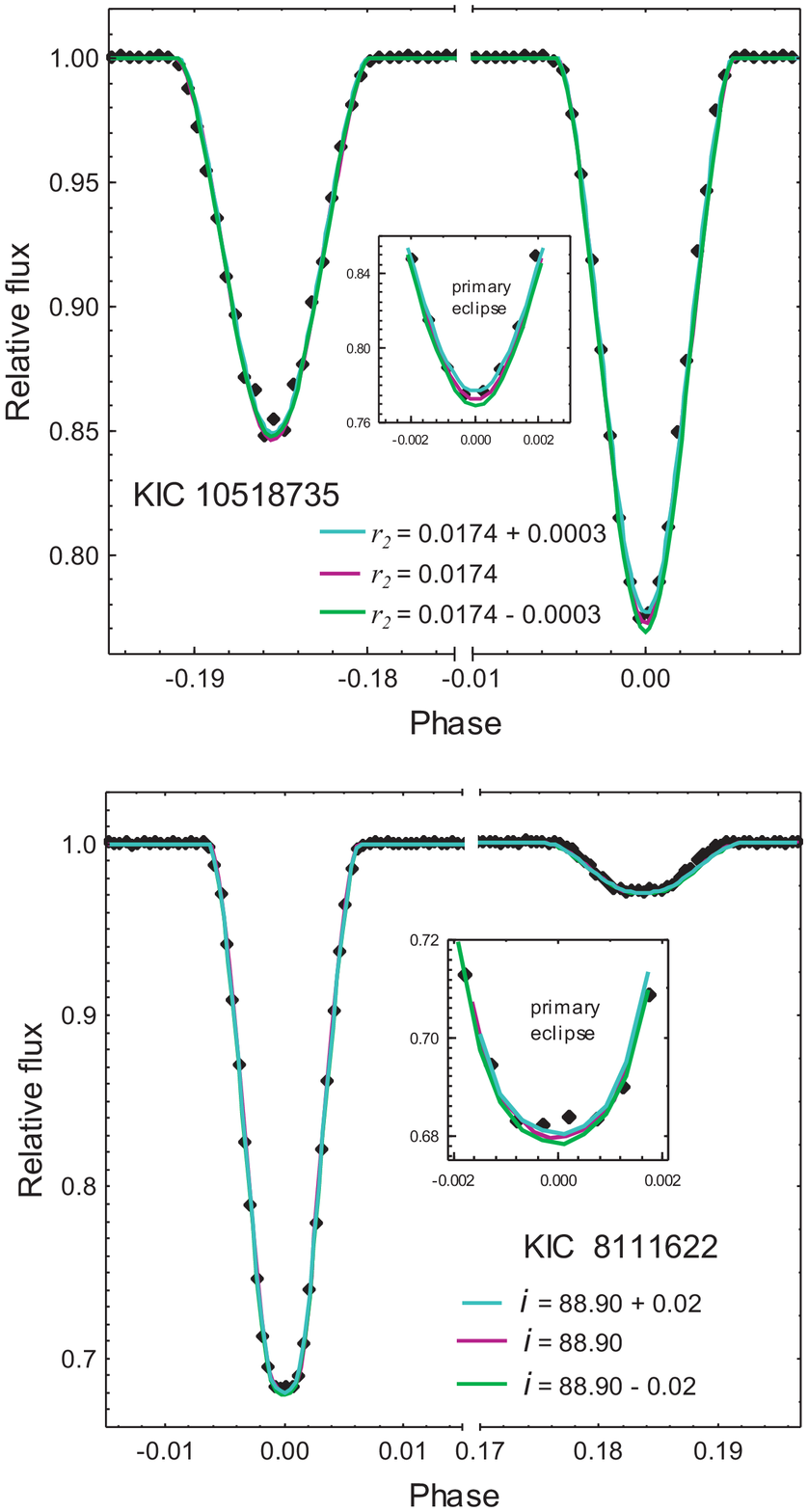}
    \caption{The three synthetic fits used for determination of the precision of fitted parameters:
    illustration for the secondary relative radius of KIC 10518735 and orbital inclination of KIC 8111622.}
   \label{Fig4}
   \end{figure}

\section{Light curve solutions}

The modeling of \emph{Kepler} data was carried out by the package
\textsc{PHOEBE} \citep{Prsa05}. We used its mode ''Detached
binaries'' because their light curves (Fig. 1) imply such
configurations.

Long cadence (LC) data from different quarters are available in
the \emph{Kepler} archive for these binaries. To ignore the effect
of accidental light fluctuations in the procedure of the light
curve solutions and to accelerate the light curve synthesis, we
modelled all available LC photometric data (above 50000 points for
each target) after phase binning. We used 2000 bins in phase (the
value of PHOEBE leading to maximum number points, important for
targets with narrow eclipses).

\begin{table}
\caption{Fixed, fitted and calculated parameters of the light
curve solutions: mass ratio $q(T^i)$ calculated by the initial
component temperatures; eccentricity $e$; periastron angle
$\omega$; inclination $i$; temperature of the secondary star
$T_{2}$; surface potentials $\Omega_{1, 2}$; relative radii of the
primary and secondary stars $r_{1, 2}$; periastron phase
$\varphi_{per}$; luminosity ratio of the stellar components
$L_2/L_1$; calculated temperatures of the primary and secondary
stars $T_{1, 2}^f$; calculated mass ratios $q(L)$ and $q(T^f)$
from the luminosity ratio and final component temperatures; value
$q_{fit}$ obtained by arbitrary varying of the mass ratio. }
\label{tab:2} \centering
\begin{tabular}{cccc}
\hline\hline
Kepler ID       &   8111622     &  10518735 &   8196180   \\
\hline
$q(T^i)$       &   0.378       &  0.858    &   0.634   \\
\hline
\emph{e}        &   0.533(2)    &  0.519(2) &   0.151(1)    \\
$\omega$ [deg]  &   165.7(3)    &  5.4(2)   &   180.5(1)    \\
\emph{i} [deg]  &   88.90(2)    &  88.42(3) &   83.40(1)    \\
$T_2$  [K]      &   3620(15)    &  4857(30) &   5319(8)    \\
$\Omega_1$      &   29.12(5)    &  40.37(9) &   8.46(2)      \\
$\Omega_2$      &   21.13(4)    &  51.05(9) &   11.53(3)  \\
\hline
$r_1$           &   0.0352(2)   &  0.0259(4)&   0.1290(3)    \\
$r_2$           &   0.0192(1)   &  0.0174(3)&   0.0622(1)   \\
$\varphi_{per}$ &   0.070       &  0.917    &   0.201      \\
$L_{2}/L_{1}$   &   0.051       &  0.36     &   0.072        \\
$T_1^f$  [K]    &   5665        &  5217     &   7246    \\
$T_2^f$  [K]    &   3681        &  4928     &   5451    \\
\hline
$q(L)$          &   0.475       &  0.774    &   0.517   \\
$q(T^f)$        &   0.502       &  0.714    &   0.634   \\
$q_{fit}$       &   0.428       &  0.764    &   0.649   \\
\hline\hline
\end{tabular}
\end{table}

\begin{figure}
   \centering
   \includegraphics[width=8.2cm, angle=0]{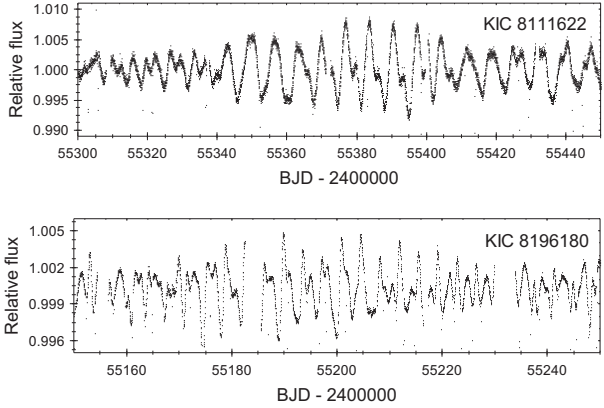}
   \caption{Out-of-eclipse variability}
   \label{Fig5}
   \end{figure}

\begin{figure}
   \centering
   \includegraphics[width=7cm, angle=0]{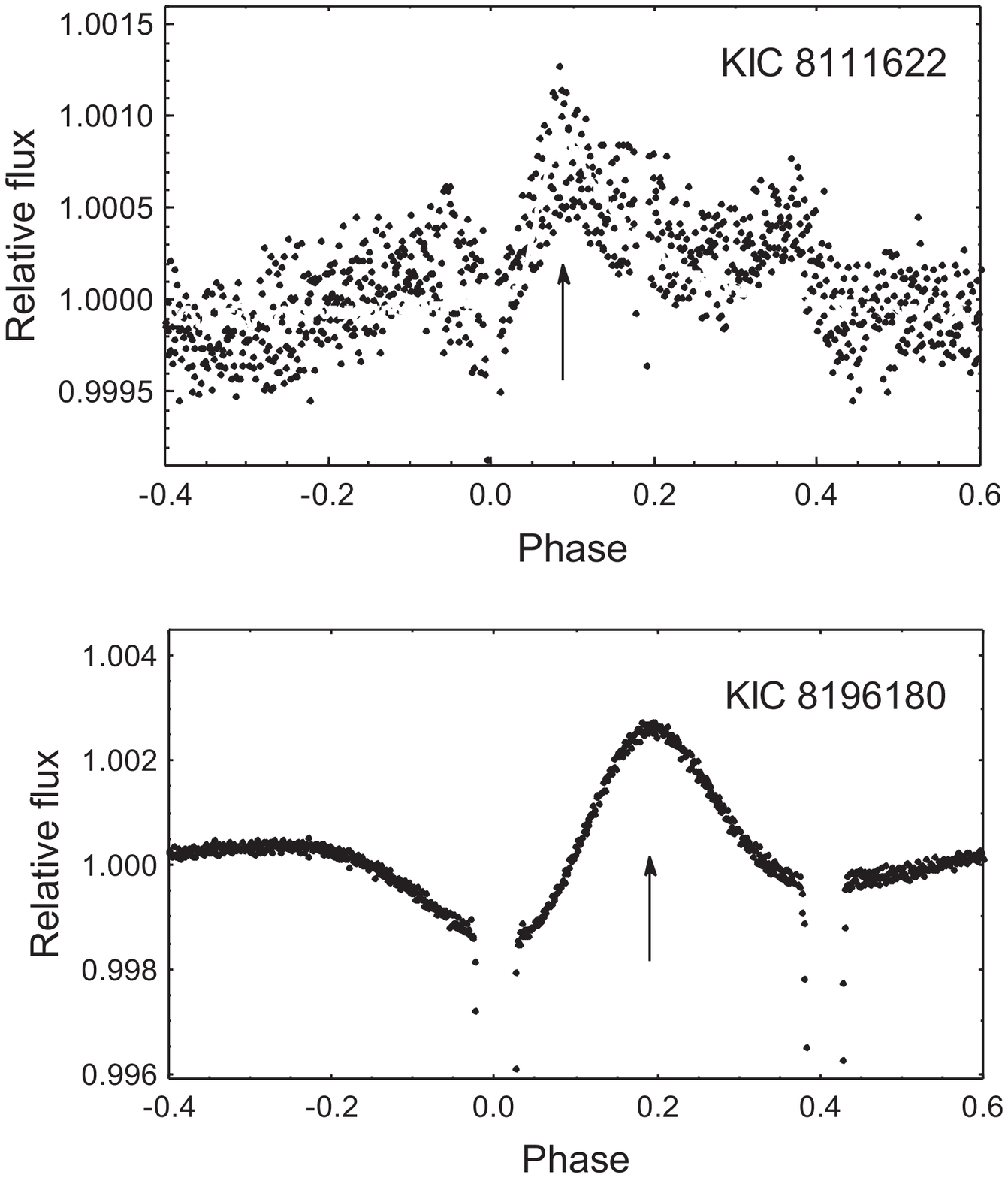}
   \caption{Tidally-induced light feature at periastron (marked by arrows).}
   \label{Fig6}
   \end{figure}

The procedure of the light curve solutions consists of several
steps.

(1) \emph{Calculation of initial (approximate) values of some
parameters }

Preliminary values of the eccentricity \emph{e} and periastron
angle $\omega$ were calculated according to the formulae
\citep{kjur15p}:
\begin{equation}
e_0 \cos\omega_0=\frac{\pi}{2}[(\varphi_2-\varphi_1)-0.5]
\end{equation}
\begin{equation}
e_0 \sin\omega_0=\frac{w_2-w_1}{w_2+w_1}
\end{equation}
where the values of $\varphi_2$, $w_1$ and $w_2$ were taken from
Table 1 ($\varphi_1$=0).

Assuming that the target components are MS stars and $T_1^i$ =
$T_{m}$, we calculated initial values of secondary temperature
$T_2^i$, mass ratio $q(T^i)$ and relative stellar radii $r_i$ by
the formulas (see Appendix)
\begin{equation}
T_2^i = T_1^i\left( \frac{d_2}{d_1}\right)^{1/4}, \quad q(T^i) =
\left( \frac{T_2^i}{T_1^i}\right)^{1.6},
\end{equation}
\begin{equation}
\frac{r_2^i}{r_1^i} = \left[q(T^i)\right]^{0.75}, \quad  r_1^i +
r_2^i \approx \pi w
\end{equation}
where $w$ is the mean eclipse width from Table 1 while $d_{1,2}$
are the depths of eclipses.

(2) \emph{Light curve solution by \textsc{PHOEBE}}

The solution of light curve for detached binary is quite
insensitive to the mass ratio in opposite to semidetached and
overcontact configurations \citep{Ter05}. That is why we fixed
this parameter of our targets (Table 2) to its value from equation
(3). To search for the best fit to the data we varied the
following parameters: \emph{e} and $\omega$ around their
preliminary values $e_0$ and $\omega_0$ and $T_{2}$ around their
initial values (3); inclination \emph{i} in the range
80-90$^{\circ}$ (appropriate for eclipsing detached stars);
potentials $\Omega_{1,2}$ around their values corresponding to the
initial values of $r_{1i}$ and $r_{2i}$ from equations (4). We
adopted coefficients of gravity brightening 0.32 and reflection
effect 0.5 appropriate for late stars (Table 1). The
limb-darkening coefficients were chosen according to the tables of
\cite{VanH93}.

After reaching the best fit (corresponding to the minimum of
$\chi^2$) we adjusted the stellar temperatures $T_{1}$ and $T_{2}$
around the value $T_m$ (Table 1) by the formulae \citep{kjur15p}

\begin{equation}
T_{1}^f = T_m + s \frac{\Delta T}{s+1},
\end{equation}
\begin{equation}
T_{2}^f = T_1 - \Delta T,
\end{equation}
where $s=L_2/L_1$ (luminosity ratio) and $\Delta T = T_m-T_2$ were
taken from the last \textsc{PHOEBE} fitting.

Although \textsc{PHOEBE} works with potentials, it gives a
possibility to calculate directly all values (polar, point, side,
and back) of the relative radius $r_i=R_i/a$ of each component
($R_i$ is linear radius and \emph{a} is orbital separation).
Moreover, \textsc{PHOEBE} yields as output parameters bolometric
magnitudes $M_{bol}^i$ of the two components in conditional units
(when radial velocity data are not available). But their
difference $M_{bol}^2-M_{bol}^1$ determines the true luminosity
ratio $s=L_2/L_1$.

The parameters of the best light curve solutions are given in
Table 2 while the corresponding synthetic curves are shown in
Figs. 1--3 as continuous lines. The residual curves show bigger
scatters during the eclipse phases (Figs. 1--3). Similar behavior
could be seen also for other \emph{Kepler} binaries \citep{ham13a,
ham13b, Lehmann13, mac14}. It was attributed to the effects of
finite integration time \citep{Kipping10}. The reasons for this
effect may be also numerical imperfectness of the physical model
\citep{Prsa16} as well as possible contribution of pulsations and
spots.

The formal \textsc{PHOEBE} errors of the fitted parameters were
unreasonably small. That is why we estimated the parameter errors
manually based on the following rule \citep{dim17}. The error of
parameter $A$ corresponds to that deviation $\Delta A$ from its
best-fit value $A^{bf}$ for which the biggest residual (usually at
the light curve minima) increases by 3$\bar{\sigma}$. The mean
photometric errors $\bar{\sigma}$ for KIC 8111622, KIC 10518735,
KIC 8196180 are correspondingly 0.0004, 0.001, 0.0001. Figure 4
illustrates our method for determination of precision of fitted
parameters.

The errors of temperatures $T_i^f$ (calculated by equations 5--6)
are around $\pm$200 K and they are due mainly to the errors of
$T_m$ (the errors of $T_2$ from PHOEBE are of order smaller, Table
2).

Besides the value $q(T^i)$ of mass ratio calculated by formula
(3), we calculated its values $q(L)$ and $q(T^f)$ by the
luminosity ratio $L_2/L_1$ from the PHOEBE solution and final
component temperatures $T_i^f$ (Table 2). For our targets the
foregoing values of the mass ratio differ by up to 25$\%$. We made
the following additional light curve solutions: (i) $q$ was fixed
to value $q(L)$; (ii) $q$ was fixed to value $q(T^f)$; (iii) $q$
was varied arbitrarily and the obtained value $q_{fit}$ is given
in Table 2. We established that the corresponding best light curve
solutions for these values were with equal values of orbital
parameters, temperatures and relative radii. Only the values of
potentials were different. This result confirmed once again the
insensitivity of the light curve solution of well detached
eclipsing binary to the mass ratio as well as the faithfulness of
the rest orbital and stellar parameters.

The analysis of the light curve solutions (Table 2) led to the
following results:

(i) KIC 8111622 and KIC 10518735 undergo total eclipses while the
eclipses of KIC 8196180 are partial;

(ii) KIC 8111622 and KIC 10518735 have eccentricity above 0.5
while that of KIC 8196180 is considerably smaller. This result is
in consistence with the expectation the shorter-period binary
(Table 1) to have the smaller eccentricity;

(iii) The components of targets are G and K stars, excluding the
primary of KIC 8196180 that is early F star;

\begin{table*}
\caption{Comparison of our values of component temperatures
and ratio of stellar radii (first three columns)
with those of Armstrong et al. (2014) (last three columns).}
\label{tab:3} \centering
\begin{scriptsize}
\begin{tabular}{ccccccccc}
\hline\hline
Kepler ID & $(T_1)_o$ & $(T_2)_o$ & $(r_2/r_1)_o$ & $(T_1)_A$ & $(T_2)_A$ & $(r_2/r_1)_A$  \\
\hline
8111622  &  5665 &  3681  & 0.55 & 5897$\pm$369  & 4587$\pm$765 & 0.62 \\
10518735 &  5217 &  4928  & 0.67 & 5216$\pm$357  & 4928$\pm$581 & 0.71 \\
8196180  &  7246 &  5451  & 0.48 & 8222$\pm$2763 & 8225$\pm$2792& 0.73 \\
\hline\hline
\end{tabular}
\end{scriptsize}
\end{table*}

(iv) The temperature difference of the components of KIC 8111622
and KIC 8196180 is considerable (1500 K and 1800 K respectively).

(v) The secondaries of all targets are almost twice smaller in
size than their primaries.

We compared (Table 3) our parameter values with those of
\cite{arm14}. The results are: (i) the primary temperatures
coincide within the errors; (ii) our values of temperatures of
secondaries of KIC 8111622 and KIC 8196180 are considerably
smaller than those of \cite{arm14}; (iii) Our values of the ratio
of stellar radii for KIC 8111622 and KIC 10518735 are quite near
to those of \cite{arm14} but the two values for KIC 8196180 are
quite different (Table 3). We checked that the synthetic light
curves corresponding to the parameter values of \cite{arm14} do
not reproduce \emph{Kepler} light curves of the targets. The
bright example is KIC 8196180 with slightly eccentric orbit and
very different eclipse depths (Fig. 3) for which \cite{arm14} have
obtained equal component temperatures (Table 3).

\section{Out-of-eclipse variability of the targets}

The out-of-eclipse light of KIC 811162 and KIC 8196180 contain
semi-regular oscillations (Fig. 5) which undergo long-term
modulation with amplitudes up to 0.01 mag. Their shape as well as
lack of strong periodicity seem to exclude pulsation explanation.
This was confirmed by the periodogram analysis of the
out-of-eclipse data that revealed bad-defined peak. The
out-of-eclipse variability of this type implies a superposition of
two periodic (or quasi-periodic) signals of similar frequency.
That produces characteristic "beating": presence of single larger
peaks followed by two smaller peaks. One may note the fast changes
of the shape and amplitude of the out-of-eclipse variability of
KIC 811162 and KIC 8196180.

\subsection{Tidally-induced effects}

Besides eclipses KIC 811162 and KIC 8196180 exhibit
tidally-induced periastron features (Fig. 6).

According to the model of \cite{Kumar95} the amplitude of the
periastron feature depends on the separation of the objects and
their masses while its shape depends on the orbit parameters
\citep{ham13b}. The contributions of different parameters on the
shape, width and amplitude of the periastron feature were
investigated by \cite{dim17}.

The shape and parameters of tidally-induced periastron features of
KIC 811162 and KIC 8196180 confirmed the theoretical predictions.

(a) The periastron feature of KIC 811162 is around 2 times
narrower than that of KIC 8196180 (Fig. 6) due to its bigger
eccentricity (fig. 4 of \cite{dim17}).

(b) The amplitude of feature of KIC 8196180 is almost 3 times
bigger than that of KIC 811162. This is due to the considerably
smaller period of KIC 8196180 and correspondingly to the bigger
relative radii of components (fig. 4 of \cite{dim17}) leading to
bigger tidal interaction. Additional reason may be its higher
temperature \citep{Kumar95}.

(c) The tidally-induced periastron features have a ''hook'' shape
consisting of two parts with almost the same widths: light
increasing preceded by light trough (that coincides with the
primary eclipse). This shape is expected for systems with $150^0
\leq \omega \leq 180^0$ (fig. 3 of \cite{dim17}). The arguments of
periastron $\omega$ of our two targets (Table 2) fulfil this
condition.

The periastron feature of KIC 811162 is \textbf{hardly} visible
(Fig. 6), partially due to its faintness (Table 1). But the clear
and well-visible tidally-induced periastron feature of KIC 8196180
means that it may be considered definitely as a heartbeat star of
EB+HB subtype \citep{dim17}.

None of our targets reveals tidally-induced pulsations.

\subsection{Spot activity}

(a) The dominant period of the out-of-eclipse variability of KIC
811162 $P_{{out}}$ = 6.728 days is not harmonic of the orbital
period $P_{orb}$ = 15.488 d. The ratio $P_{orb}/P_{out} \sim$ 2.3
is near, but differs from the ratio of pseudo-synchronous angular
velocity and mean motion $\Omega_{ps}/n \sim$ 2.74 corresponding
to $e$ = 0.53 of \textbf{KIC 811162} \citep{Hut81}.

The amplitude of periastron feature of KIC 811162 is around 0.0007
in relative fluxes (Fig. 6) while the amplitudes of the total
out-of-eclipse variations are up to 0.0015 (Fig. 5). This means
that the amplitudes of the additional variations do not exceed
0.0008 (in relative flux). Probably, $P_{{out}}$ presents
rotational period of some of the target components and the
observed additional out-of-eclipse variability (besides periastron
feature) is due to variable spot visibility. Then the bad-defined
peak of the Fourier transform may be considered as a result of
differential rotation of the asynchronous component while the
long-term modulation may be attributed to stellar activity cycles.
The asynchronism of KIC 811162 is expected considering its high
eccentricity and long orbital period.

The cycles of out-of-eclipse variability of KIC 811162 alternate
between bigger-amplitude single-peaked shape and smaller-amplitude
two-peaked shape (Fig. 5). Two spots on almost opposite longitudes
of the asynchronous component could explain the two-peaked cycles.
However, the single-peaked cycles cannot be reproduced by one spot
if the highly-inclined configuration of KIC 811162 is coplanar (it
would cause light variation whose light curve is flat during
almost half a cycle, different from the observed one in Fig. 5).
There are two possible alternatives: KIC 811162 is not a coplanar
binary (expected for asynchronous system) with arbitrary spot
location or KIC 811162 is coplanar binary with big almost polar
spot.

(b) The amplitude of periastron feature of KIC 8196180 is around
0.004 (in relative fluxes) while the amplitudes of the total
out-of-eclipse variations are up to 0.008. This means that the
amplitudes of the additional variations do not exceed 0.004. The
dominant period of the total out-of-eclipse variability is the
orbital one that means the additional variability (besides the
periastron feature) to be also with the orbital period. The
two-peaked shape of the variations (Fig. 5) and their period imply
rotational variability of synchronously-rotating component with
two spots on almost opposite longitudes. The variable amplitudes
and durations of the two waves per cycle may due to differential
rotation and variable latitude of the spots.

\section{Conclusions}

This paper presents the results of determination of the orbits and
parameters of stellar configurations of the eclipsing eccentric
binaries KIC 8111622, KIC 10518735 and KIC 8196180 on the basis of
their \emph{Kepler} data. KIC 8196180 turned out heartbeat star.
The characteristics of its tidally-induced feature are consistent
with the theoretical predictions. Moreover, KIC 8196180 exhibits
out-of-eclipse light variations with the orbital period that may
be attributed to spot(s) on synchronously rotating component.
Besides \textbf{hardly} visible periastron feature KIC 811162
exhibits small-amplitude light variations whose period is around
2.3 times shorter than the orbital one. They could be attributed
to spots on asynchronously rotating component. The asynchonism of
KIC 8111622 was expected taking into account its long orbital
period.

The presented study adds new member to the family of heartbeat
binaries and provides new data to search for dependencies of the
tidally-induced effects of eccentric binaries on their orbital and
stellar parameters.

\acknowledgments This work was supported partly by project DN08/20
of the Fund for Scientific Research of the Bulgarian Ministry of
Education and Science and by project RD 08-102/03.02.17 of Shumen
University. It used the SIMBAD database and NASA Astrophysics Data
System Abstract Service. We used data from the \emph{Kepler} EB
catalog (http://keplerebs.villanova.edu/). The authors are very
grateful to the anonymous referee for the valuable notes and
recommendations.


\section{Appendix}

1) Within the black-body approximation the relative fluxes at the
bottom of the two eclipses are

\begin{equation}
f_1 = \frac{\pi k_1^{2}T_1^{4}+\pi k_2^{2}T_2^{4} - \pi k_2^{2}T_1^{4}}{\pi k_1^{2}T_1^{4}+\pi k_2^{2}T_2^{4}}
\end{equation}

\begin{equation}
f_2 = \frac{\pi k_1^{2}T_1^{4}}{\pi k_1^{2}T_1^{4}+\pi k_2^{2}T_2^{4}}
\end{equation}
where $k_i=R_i/d$, $R_i$ is the stellar radius, \emph{d} is target
distance).

By introducing $X=(r_2/r_1)^2$, $Y=(T_2/T_1)^4$ and eclipse depths
$d_1=1-f_1$ and $d_2=1-f_2$ one can obtain
\begin{equation}
X = \frac{d_1}{1-d_2}, \quad  Y = \frac{d_2}{d_1},
\end{equation}
that means
\begin{equation}
T_2/T_1 = \left( \frac{d_2}{d_1}\right)^{1/4} .
\end{equation}

The expressions (7--8) refer to totally-eclipsed binary. For
partial eclipses the corresponding expressions are

\begin{equation}
f_1 = \frac{\pi k_1^{2}T_1^{4}+\pi k_2^{2}T_1^{4} - a \pi
k_2^{2}T_1^{4}}{\pi k_1^{2}T_1^{4}+\pi k_2^{2}T_2^{4}}
\end{equation}

\begin{equation}
f_2 = \frac{\pi k_1^{2}T_1^{4}+\pi k_2^{2}T_2^{4} - a \pi
k_2^{2}T_2^{4}}{\pi k_1^{2}T_1^{4}+\pi k_2^{2}T_2^{4}}
\end{equation}
where $a$ is the relative area of the secondary component covered
at secondary eclipse ($0 < a <1$). One can check that the
expressions (11--12) lead to the same relation (10).

2) If one assumes that the target components are MS stars then
empirical relations between stellar parameters are applicable.
From $L \sim M^4$ and $R \sim M^{3/4}$ one can obtain
\begin{equation}
q=M_2/M_1 = (L_2/L_1)^{1/4} = [(R_2/R_1)^2 (T_2/T_1)^4]^{1/4},
\end{equation}
i.e.
\begin{equation}
q = (M_2/M_1)^{(3/4)(1/2)} (T_2/T_1) = q^{3/8} (T_2/T_1) .
\end{equation}
As a result
\begin{equation}
q = (T_2/T_1)^{8/5}
\end{equation}
and
\begin{equation}
r_2/r_1 = R_2/R_1 = q^{3/4} .
\end{equation}

3) The primary eclipse ends at phase $\varphi_e$ of outer contact
of the stellar components that can be obtained from the approximate
formula
\begin{equation}
r_1 + r_2 \approx \sin (2\pi \varphi_e) .
\end{equation}
For well detached binaries with narrow eclipses one may substitute
$\sin (2\pi\varphi_e)$ with $2\pi\varphi_e$. Thus, we
obtained
\begin{equation}
r_1+r_2 = \pi w
\end{equation}
where $w$ is the eclipse width in phase units.

Part of the considerations in the Appendix can be found in
\cite{kjur06bg}, \cite{iva10} and \cite{kjur17aj}.

\end{document}